\documentclass[11pt]{article}
\usepackage{amsmath,amssymb,dsfont}
\newcommand{\ket}[1]{\left|#1\rangle\right.}
\newcommand{\beq}{\begin{equation}}
\newcommand{\eeq}{\end{equation}}
\newcommand{\ba}{\begin{array}}
\newcommand{\ea}{\end{array}}
\newcommand{\bea}{\begin{eqnarray}}
\newcommand{\eea}{\end{eqnarray}}
\newcommand{\bean}{\begin{eqnarray*}}
\newcommand{\eean}{\end{eqnarray*}}

\newtheorem{theorem}{Theorem}[section]
\newtheorem{prop}[theorem]{Proposition}

\newtheorem{remark}[theorem]{Remark}
\newenvironment{rem}{\begin{remark} \rm}{\end{remark}}

\newtheorem{proof}{Proof.}

\newcounter{appendix}
\newcommand{\CH}{{\cal H}}

\newcommand{\CD}{{\cal D}}

\newcommand{\CG}{{\cal G}}

\newcommand{\CU}{{\cal U}}

\def\l{\lambda}

\def\binomial#1#2{{#1\choose #2}}
\newcommand{\half}{\frac{1}{2}}

\def\mat2#1#2#3#4{{\left(\begin{array}{cc}#1 & #2\\ #3 & #4
      \end{array}\right)}}

\def\mats2#1#2#3#4{{\left(\begin{array}{cc}#1 & #2\vspace{2truemm} \\ #3 & #4
\end{array}\right)}}

\def\endpf{\begin{flushright}$\square$\end{flushright}}

\begin{document}
\begin{center}
{\huge Gaudin models with ${\CU}_q(\mathfrak{osp}(1 | 2))$ symmetry \\}
\end{center}
\vspace{0.3truecm}
\begin{center}
{\large
Fabio Musso$^\diamondsuit$, 
Matteo Petrera$^\sharp$, Orlando Ragnisco$^\flat$, Giovanni Satta$^\natural$}
\vskip0.3truecm
Dipartimento di Fisica E. Amaldi \\
Universit\`a degli Studi di Roma Tre\\
and\\
Istituto Nazionale di Fisica Nucleare,
Sezione di Roma Tre\\
Via della Vasca Navale 84, 00146 Roma, Italy\\
\vspace{0.3truecm}
E--mail$^\diamondsuit$:  musso@fis.uniroma3.it\\
E--mail$^\sharp$:  petrera@fis.uniroma3.it\\
E--mail$^\flat$:  ragnisco@fis.uniroma3.it\\
E--mail$^\natural$:  satta@fis.uniroma3.it\\
\end{center}
\vspace{0.2truecm}
\abstract{\noindent
We consider a Gaudin model related to the $q$-deformed
superalgebra ${\CU}_q(\mathfrak{osp}(1 | 2))$. 
We present an exact solution to that system diagonalizing a complete set of commuting observables, 
and providing the corresponding eigenvectors and eigenvalues. The
approach used in this paper is based on the coalgebra supersymmetry of the model.}
\vspace{1truecm}

\section{Introduction}

The Gaudin model, introduced by M. Gaudin in 1976, is a quantum mechanical system 
involving long--range spin interaction \cite{G1,G2}.

In \cite{SK1} it was solved in the framework of the algebraic Bethe Ansatz. 
It was also shown there that the model is governed by a Yang--Baxter algebra, called 
the Gaudin algebra, with commutation relations linear in the generators and determined 
by a classical $r$-matrix. It is to be stressed that this features are present in the model
despite its quantum mechanical nature. In fact the Gaudin model is one of a large class 
of models, with such an algebraic nature, so that its study becomes an important issue.

Let us recall that the superalgebra extension of the Gaudin algebra, and of the 
related $r-$matrix structure, has been worked out in some remarkable 
papers (see for instance \cite{MC,KU}) where the Gaudin 
model related to orthosymplectic Lie superalgebra $\mathfrak{osp}(1|2)$ has been 
constructed  and solved through  a brilliant generalization of the 
Bethe-Ansatz.

It is known that this algebraic richness and robustness allows one to
use it as a testing  ground for many ideas such as the Bethe Ansatz and the general procedure
of separation of variables. 

Among these approaches, the coalgebraic one was introduced in a series of papers 
\cite{BCR,BR,MR,MR2}.
A general and constructive connection between coalgebras and integrability
can be stated as follows: given {\it any} coalgebra $(\mathfrak{g},\Delta)$ with
Casimir element $C$, each of its representations gives rise to a family of
completely integrable Hamiltonians $H^{(m)}$, $m=1,...,N$ with an arbitrary number $N$
of degrees of freedom. 

Endowing this coalgebra with a suitable additional structure (either a
Poisson bracket or a non--commutative product on $\mathfrak{g}$), both classical and quantum mechanical
systems can be obtained from the same $(\mathfrak{g},\Delta)$.
It is important to emphasize that the validity of this general procedure by
no means depends on the explicit form of $\Delta$ (i.e., on whether the
coalgebra $(\mathfrak{g},\Delta)$ is deformed or not).

In this framework 
a particular class of coalgebras that can be used to
construct systematically integrable systems are the so--called {\it q-algebras} \cite{TJ}.
The feature of
such systems will be that they are integrable deformations of the
ones obtained applying the same method to the corresponding
non-deformed coalgebra.

We briefly recall here a general construction of completely integrable quantum systems
associated with Lie (rank-1) superalgebras based on a coalgebraic approach \cite{J}.
Applying the method to higher ranks (super)algebras it is still possible to obtain
commuting observables but completeness is by no means guaranteed \cite{PIP}.
\vspace{.4truecm}

Let us consider a Lie superalgebra $\mathfrak{g}$ with Casimir $C \in \CU(\mathfrak{g})$,
and a co-associative 
linear mapping $\Delta : \CU(\mathfrak{g}) \mapsto \CU(\mathfrak{g}) \, \otimes \, \CU(\mathfrak{g})$ 
(denoted as coproduct) such that 
$\Delta$ is a Lie homomorphism:
$$ 
[\Delta (a),\Delta (b)\}= 
\Delta ([a,b\}), \qquad  \forall \, a,b \in \CU(\mathfrak{g}),
$$
where $[\cdot,\cdot\}$ denotes the supercommutator.
The coassociativity property allows one to
construct from $\Delta$ in an unambiguous way subsequent homomorphisms
$$
\Delta^{(2)}\doteq \Delta, \qquad 
\Delta^{(3)}:\CU(\mathfrak{g})\,  \mapsto \,\CU(\mathfrak{g})^{\otimes}{}^3, \quad \dots \quad,
\Delta^{(N)}:\CU(\mathfrak{g})\,  \mapsto \,\CU(\mathfrak{g})^{\otimes}{}^N.
$$
Thus, we can associate to our superalgebra, (or better {\it{co-superalgebra}}) a
quantum integrable system with $N$  degrees of freedom,
whose Hamiltonian is an arbitrary function of the $N$-th
coproduct of the generators and the remaining $N-1$ integrals of motion are provided by
$\Delta^{(m)} (C)$, $m=2,\dots ,N$.

In \cite{ACF,BGT} it has been shown how to associate to a 
Lie--Hopf superalgebra a quantum integrable system and how to extend
this procedure to $q$-superalgebras. In fact, $q$-superalgebras are obtained by
Lie--Hopf superalgebras through a process of deformation that preserve their
Lie--Hopf structure. It is therefore possible to associate to $q$-superalgebras
integrable systems that are deformed version of the ones associated
to the original superalgebra.

In the present paper we will consider an integrable $q$-deformation of the Lie superalgebra
$\mathfrak{osp}(1 | 2)$ in order to obtain a Gaudin model with 
${\CU}(\mathfrak{osp}(1 | 2))$ symmetry.

\section{A $q$-deformation of ${\CU}(\mathfrak{osp}(1 | 2))$}

The quantum superalgebra ${\CU}_q(\mathfrak{osp}(1 | 2))$ \cite{SA,PI}
as a deformation of the universal enveloping algebra
of the Lie superalgebra $\mathfrak{osp}(1 | 2)$ is generated by three elements $E,F,H$. The $q$-deformed
commutation relations between the generators are:
\beq
\left\{E,F\right\}= \frac{q^H-q^{-H}}{q-q^{-1}}, \qquad
[H,E]=E, \qquad
[H,F]=-F\label{comm}
\eeq
In the following we will also need the operators $F^2$ and $E^2$ fulfilling the commutation
relations
\begin{eqnarray*}
&& \left[F^2,E\right] = \kappa \left( q^{H+1/2}+q^{-H- 1/2}  \right) F, \\
&& \left[E^2,F\right] = -\kappa \left(q^{H-1/2}+q^{-H+1/2} \right) E, \\
&& \left[E^2,F^2\right]= \kappa^2 - \kappa \frac{q^{2H+1/2}-q^{-2H-1/2}}{q-q^{-1}} +
(q^H-q^{-H}) E\,F, \\
&& [H, E^2 ]= 2 \, E^2, \\
&& [H, F^2 ]= -2 \, F^2,
\end{eqnarray*}
where
$$
\kappa \doteq  \frac{1}{q^{1/2}+q^{-1/2}}.
$$
The center of ${\CU}_q(\mathfrak{osp}(1 | 2))$ is spanned by the $q$-deformed Casimir element
(provided that $q$ is not a root of the unity \cite{AR}):
\beq
C(q)= \left( \frac{q^{H-1/2}+q^{-H+1/2}}{q-q^{-1}} \right)^2 
- \kappa^2 E^2F^2 - \left( q^{H-1}-q^{-H+1}\right) E\,F . \label{cs}
\eeq
We now endow ${\CU}_q(\mathfrak{osp}(1 | 2))$ with a coalgebra structure. This can be done
assigning the following $q$-deformed coproduct:
\bea
&& \Delta_q(H) = H\otimes \mathds{1}+\mathds{1}\otimes H, \nonumber \\
&& \Delta_q(E) = E\otimes q^{\frac{H}{2}}+q^{-\frac{H}{2}}\otimes E, \label{cazzetti} \\
&& \Delta_q(F) = F\otimes q^{\frac{H}{2}}+q^{-\frac{H}{2}}\otimes F. \nonumber
\eea
which establish a superalgebra homomorphism:
\begin{eqnarray*}
&& \left\{\Delta_q(E),\Delta_q(F)\right\}=\frac{\Delta_q(q^H)-\Delta_q(q^{-H})}{q-q^{-1}},\\
&& \left[\Delta_q(H),\Delta_q(E)\right]=\Delta_q(E),\\
&& \left[\Delta_q(H),\Delta_q(F)\right]=-\Delta_q(F).
\end{eqnarray*}

For the sake of completeness we give the corresponding antipode and counit,
$$
\epsilon(H)=\epsilon(E)= \epsilon(F)=0, \qquad \epsilon(q^{\pm H}) = 1,
$$
$$
\sigma(E)= -q\,E,\qquad 
\sigma(F)= -q^{-1}F,\qquad 
\sigma(H)= -H,\qquad 
\sigma(q^{\pm H})=q^{\mp H},
$$
obtaining a Hopf superalgebra.

The coproducts (\ref{cazzetti}) can be extended to the $N$-th order by means of the
coassociativity property as in the non deformed case, taking into account that
$$
\Delta_q (q^H)= q^H \otimes q^H.
$$
Explicitly,
\bea
&& \Delta^{(N)}_q(H)= \sum_{i=1}^N H_i, \nonumber \\
&& \Delta^{(N)}_q(E)= \sum_{i=1}^N  E_i\, q^{\frac{1}{2}\sum_{j=1}^N {\rm{sgn}}(i-j) H_j}, \nonumber \\
&& \Delta^{(N)}_q(F)= \sum_{i=1}^N  F_i\, q^{\frac{1}{2}\sum_{j=1}^N  {\rm{sgn}} (i-j) H_j}. \nonumber 
\eea

\begin{rem} In the limit $q \rightarrow 1$ the above deformed supercommutation relations
obviously reduce to well-known supercommutation relations of the Lie superalgebra
$\mathfrak{osp}(1 | 2)$ \cite{FR}. Let us recall that $\mathfrak{osp}(1 | 2)$ has dimension five and rank
one; the supercommutation relations between its generators are
$$
\left\{E,F\right\}= H, \qquad
[H,E] = E, \qquad
[H,F] = F.
$$
$$
\left\{E,E\right\}= 2 E^2, \qquad 
\left\{F,F\right\}= 2 F^2, \qquad 
$$
$$
[E^2, F]= -E, \qquad [F^2, E]= F,
$$
$$
[H, E^2] = 2E^2, \qquad 
[H, F^2]= -2 F^2, \qquad
[F^2, E^2] =H.
$$
We see that the operators $H,E^2,F^2$ generate the Lie algebra $\mathfrak{sl}(2)$. 
The above supercommutation relations define  $ H, E^2,F^2 $ as the  bosonic generators, and 
$ E,F$ as the fermionic ones, i.e. ${\rm{deg}}(H)={\rm{deg}}(E^2)={\rm{deg}}(F^2)=0$ and
${\rm{deg}}(F)={\rm{deg}}(E)=1$.
This gradation can naturally be extended to the deformed enveloping superalgebra 
${\CU}_q(\mathfrak{osp}(1 | 2))$, since 
\beq \label{bg2}
{\rm{deg}}(a\,b)= {\rm{deg}}(a) + {\rm{deg}}(b),\qquad {\rm{mod}}\,2 \qquad
\forall \, a,b \in {\CU}_q(\mathfrak{osp}(1 | 2)),
\eeq
and ${\rm{deg}}(q^H)=1$.

In the same limit $ q \rightarrow 1 $, definitions (\ref{cazzetti}) also reduce to the non--deformed 
coproducts for the superalgebra  $\mathfrak{osp}(1 | 2)$, wich we will denote with $\Delta $.

\end{rem}

\begin{rem} In order to obtain a superalgebra homomorphism from the coproduct
(in the deformed case $ \Delta_q $ as in the non--deformed one) a necessary 
requirement is that the tensor product be a suitable graded one. The proper
definition of moltiplication between $N$ elements tensor products is the following one \cite{J}:
\beq
(a_1 \otimes \dots \otimes a_N)(b_1 \otimes \dots \otimes b_N)=
(-1)^{\sum_{i < j=2}^N {\rm{deg}}(a_j){\rm{deg}}(b_i)}
(a_1 \, b_1) \otimes \dots \otimes (a_N \, b_N), \label{bg}
\eeq
for all $a_j,b_i \in {\CU}_q(\mathfrak{osp}(1 | 2))$. Notice that (\ref{bg2}), together with the
definitions (\ref{cazzetti}), assures
that
$$
{\rm{deg}} (\Delta_q^{(m)}(a)) = {\rm{deg}} (a), \qquad \forall a \in {\CU}_q(\mathfrak{osp}(1 | 2)),
\, m \in \mathbb{N}.
$$
In other words the deformed coproduct preserves the gradation of the superalgebra.
\end{rem}

\section{Exact solution of a ${\CU}_q(\mathfrak{osp}(1 | 2))$ Gaudin model}

Now we have all we need to construct a Gaudin model with ${\CU}_q(\mathfrak{osp}(1 | 2))$ symmetry in 
the coalgebra setting.
 
We consider
the  $N$ commuting observables $\{C^{(n)}(q)\}_{n=1}^N$:
$$
\left[C^{(m)}(q),C^{(n)}(q)\right]=0, \quad \forall \, m,n=1,...,N,
$$
where
\bea
C^{(m)}(q) &=& \Delta^{(m)}_q\left[C(q)\right] = \nonumber\\
&=&\left[ \frac{\Delta^{(m)}_q\left(q^{H-1/2}\right)+\Delta^{(m)}_q\left(q^{-H+1/2}\right)
}{q-q^{-1}}
 \right]^2 - \kappa^2 \Delta^{(m)}_q( E^2)\,\Delta^{(m)}_q(F^2)+ \nonumber \\
&& \nonumber \\
&& - \left[\Delta^{(m)}_q\left( q^{H-1}\right)-\Delta^{(m)}_q
\left(q^{-H+1}\right)\right]\Delta^{(m)}_ q(E)\,\Delta^{(m)}_q(F) .
\nonumber
\eea
Hereafter we parametrize the deformation parameter with $ z \doteq \ln q $.

A ``physical'' Gaudin Hamiltonian for the $N$-bodies system can be choosen as the $N$-th order deformed 
coproduct of 
the  Casimir $\Delta_q^{(N)}\left[C(z)\right]$, namely
\bea \label{ham}
&& \CH_q =
\frac{\mbox{sinh}^2\left[z\left(\Delta_q^{(N)}(H)-\frac{1}{2}\right)\right]}{\mbox{sinh}^2z}
-\kappa^2\Delta_q^{(N)}(E^2)\,\Delta_q^{(N)}(F^2)+ \nonumber\\
&&{} \nonumber\\
&& \qquad -2\,\mbox{cosh}\left[z\left(\Delta_q^{(N)}(H)-1\right)\right]\Delta_q^{(N)}(E)\,\Delta_q^{(N)}(F).
\eea

This Hamiltonian can be written in any representation of the deformed superalgebra  ${\CU}_q(\mathfrak{osp}(1 | 2))$.
While it's always possible to choose a particular one, we will work in the general 
case of spin $j$ representation
(with integer or half-integer $j$). Further generalization can be obtained by allowing site--dependent
representations $(j_1, ..., j_N)$. However, for the sake of simplicity,
we will not deal with this more general case in the present paper.

A complete set of independent commuting observables is provided by
\beq \label{oss}
\left\{\Delta_q^{(N)}(H),C^{(2)}(z),\dots,C^{(N)}(z)\right\}.
\eeq
We can write the Hamiltonian (\ref{ham}) in the following form:
\beq
\CH=
\frac{\mbox{sinh}^2\left[z\left(\sum_{i=1}^N H_i -\frac{1}{2}\right)\right]}{\mbox{sinh}^2z}
-\kappa^2\sum_{i,j,k,l=1}^N\eta_i\,\eta_j\,\phi_k\,\phi_l
-2\,\mbox{cosh}\left[z\left(\sum_{i=1}^N H_i-1\right)\right]\sum_{i,j=1}^N\eta_i\,\phi_j, \nonumber
\eeq
where
$$
\eta_i \doteq E_i\, q^{\frac{1}{2}\sum_{j=1}^N {\rm{sgn}}(i-j) H_j},\qquad \phi_i\doteq F_i\, 
q^{\frac{1}{2}\sum_{j=1}^N  {\rm{sgn}} (i-j) H_j}.
$$
Notice that the interaction involves more than two sites: this non--local feature is a peculiar property 
of $q$--deformed models.

We will show that the common eigenstates of the family of observables (\ref{oss}) take the form
\beq
\varphi_z(k,m_l,s_{m_l}; \dots, 0,0)=
\left[ \Delta_q^{(N)}(E) \right]^{k-m_l}\psi_z(m_l,s_{m_l};\dots, 0,0), \label{stt}
\eeq
where $\psi_z(m_l,s_{m_l};\dots;0,0)$ is an element of the basis spanning the kernel of the lowering
 operator $\Delta_q^{(s_{m_l})}(F)$. These elements can be obtained through the recursive formula:
\beq 
\psi_z(m_l,s_{m_l};\dots;0,0)=
\sum_{i=0}^{\delta m}\alpha_i(z)
\left[\Delta_q^{(s_{m_l}-1)}(E)\right]^{\delta m-i}(E_{s_{m_l}})^i\,
\psi_z(m_{l-1},s_{m_{l-1}};\dots;0,0), \label{st}
\eeq
where $\delta m \doteq m_l-m_{l-1}$ and $\{\alpha_i(z)\}_{i=1}^{\delta m}$ 
denotes a set of suitable coefficients.
Since in each representation of  ${\CU}_q(\mathfrak{osp}(1 | 2))$ we have $E^{4j+1}=0$, $j$ 
being the spin of the
choosen representation,
the sum in  formula (\ref{st})  will have at most $4j+1$ terms, so that $ \delta m \leq 4j$. 
If we consider the pseudo--vacuum state
\beq
\psi_z(0,0)= | \downarrow \cdots \downarrow \rangle \in {\rm{Ker}}\left(\Delta_q^{(k)}(F)\right), \qquad 
\forall \,
k=1, \dots, N, \label{vuoto}
\eeq
we recognize that $m_l$ stands for the total number of excitations with respect to 
 $\psi_z(0,0)$ and $s_{m_l}$ indicates the number of the last excited site (counting from the left).

\begin{prop}\label{mostro} The states (\ref{st}) are annihilated by $\Delta_q^{(s_{m_l})}(F)$ iff
\beq
\frac{\alpha_{i+1}(z)}{\alpha_i(z)}  = (-1)^{i+1} e^{\frac{z}{2}(\tau+\delta m -2)}
\frac{(-1)^{\delta m - i}\sinh\left[z\left(\tau+ \delta m -i- \half \right)\right]-
\sinh\left[z\left(\tau-\half\right)\right]}
{(-1)^{2j} \sinh \left[ z \left(j + \half \right)\right] +
\sinh \left[ z \left(i -j +\half \right)\right] }, 
\eeq 
$i=0,...,\delta m -1$, 
where $\tau$ is the eigenvalue of $\Delta_q^{(s_{m_l})} (H)$.
\end{prop}

{\bf{Proof:}} A straightforward computation. Notice that it may be useful 
the following expression
$$
F\,E^k+(-1)^{k-1}E^kF=
\frac{(-1)^{k-1} E^{k-1}}{2 \sinh z \cosh \frac{z}{2}}
\left\{ (-1)^{k} \cosh \left[z\left( H + \half \right)   \right] +
\cosh \left[z\left( H - \half \right)   \right] \right\},
$$
holding for all $k \in \mathbb{N}$. The above formula is a plain consequence of the
supercommutation relations (\ref{comm}).

\endpf

Up to a normalization constant, proposition \ref{mostro} determines all coefficients 
$\alpha_i(z)$ with $i=1,...,\delta m$.

\begin{prop}\label{mostro2} The states (\ref{stt}) are eigenvectors of the set (\ref{oss}), namely
\beq
C^{(n)}(z)\, \varphi_z(k,m_l,s_{m_l}, \dots, 0,0)
=\l_{n}(z)\, \varphi_z(k,m_l,s_{m_l}, \dots,0,0),\label{1}
\eeq
with eigenvalues $\l_{n}$ given by
\beq
\l_{n}=
\frac{\sinh^2 \left[
z\left(\rho_z -\frac{1}{2}\right)\right]}{\sinh^2z}, \label{2}
\eeq
where $\rho_z$ is the eigenvalue of $\Delta_q^{(n)}(H)$ on the state $\psi_z(i,s_i,\dots)$,
and the value of $i \leq l$ is selected by the condition
\beq
s_{m_i} \leq n < s_{m_{i+i}}, \qquad s_{m_{l+1}}=N+1. \label{3333}
\eeq
\end{prop} 

{\bf{Proof:}} Notice that
$$
C^{(n)}(z)\,\varphi_z(k,m_l,s_{m_l}, \dots, 0,0)= 
\left[ \Delta_q^{(N)}(E) \right]^{k-m_l} C_h(z)\,\psi(m_l,s_{m_l}, \dots, 0,0),
$$
since $[C^{(h)}(z), \Delta_q^{(N)}(E)]=0$ for all $n=1,...,N$. 

\noindent If
$n \geq s_{m_l}$ we readily get (\ref{1}--\ref{2}) since
$\psi_z(m_l,s_{m_l}, \dots, 0,0)$ is in  ${\rm{Ker}}\, \Delta_q^{(n)}(F)$.
Otherwise, if $n < s_{m_l}$ we can note that
$$
\left[
C^{(n)}(z), \sum_{i=0}^{\delta m}\alpha_i(z)
\left[\Delta_q^{(s_{m_l}-1)}(E)\right]^{\delta m-i}(E_{s_{m_l}})^i
\right]=0,
$$
so that we can act with $C^{(n)}(z)$ on $\psi_z(m_{l-1},s_{m_{l-1}}, \dots,0,0)$. By iteration
we will arrive to a value of $i$ such that condition (\ref{3333}) holds and to
a function $\psi_z(i,s_i,\dots)$ which fixes the value of $\rho_z$ and so the eigenvalue (\ref{2}). 
This proves the proposition.

\endpf

\begin{rem} We stress the fact that our approach has a simple algebraic interpretation. Indeed, each 
eigenstate $\varphi_z(k,m_l,s_{m_l}, \dots, 0,0)$ has  to be a basis vector
of the tensor product representation
\beq \label{degen}
\left[ \CD^{(j)}  \right]^{\otimes N} = \bigoplus_{l=0}^{Nj}
c_{j,l}^{(N)}\,  \CD_l, 
\eeq
where $\CD^{(j)}$ denotes the representation of each site and 
$\{c_{j,l}^{(N)}\}$ is the set of Clebsch--Gordan coefficients. 
Our method constructs first the lowest weight vectors $\psi_z(m_l,s_{m_l}, \dots, 0,0)$
for each $\CD_l$, taking account that $l= Nj-m_l$ and then allows  us to complete the basis with 
suitable raising operators.

Thanks to the Schur's Lemma the eigenvalues of the family (\ref{oss}) are the values
taken by the Casimir (\ref{cs}) on each $\CD_l$. Furthermore the coefficients 
$\{c_{j,l}^{(N)}\}$ are related to  the spectrum degeneracies; in fact 
the number of eigenstates of the Hamiltonian (\ref{ham}) that belong
to the
the energy eigenvalue 
corresponding to the representation $\CD_l $ is given by
$$
g_{j,l}^{(N)}=c_{j,l}^{(N)}(4l+1),
$$
being the factor $4l+1$ the degeneracy of each $\CD_l$. This latter term
could be removed by an external field, while the first one remains.
\end{rem}

\begin{rem} This graded model shares a remarkable feature with other supersymmetric
integrable systems \cite{SEE}. Namely, as we show in Appendix 1,
it is possible to assign an arbitrary grading to the pseudo--vacuum state (\ref{vuoto}).
Each choice give rise, through the above construction, to a different family of eigenstates
although the spectrum remains the same. In Appendix 2 we explicitly present two
families of eigenstates for the ${\CU}_q(\mathfrak{osp}(1 | 2))$ Gaudin model with $j=1/2$ and
$N=2$.
\end{rem}

\subsection{The $q \rightarrow 1$ limit}

We now obtain some known results \cite{J} on the Gaudin model with $\mathfrak{osp}(1 | 2)$ symmetry
considering the limit $q \rightarrow 1$. 

The family of $N$ commuting observables is $\{C^{(n)}\}_{n=1}^N$:
$$
\left[C^{(m)},C^{(n)}\right]=0, \quad \forall \, m,n=1,...,N,
$$
where
\beq
C^{(m)} =\Delta^{(m)}(C) = 
\left[\Delta^{(m)}(H)\right]^2-2\left\{\Delta^{(m)}(E^2),\Delta^{(m)}(F^2)\right\}
- \left[\Delta^{(m)}(E),\Delta^{(m)}(F)\right]. \nonumber 
\eeq
A ``physical`` non--deformed Gaudin Hamiltonian for the 
$N$-bodies system can be choosen as the $N$-th order 
coproduct of the  Casimir $\Delta^{(N)}\left(C \right)$, namely
\beq \label{ano}
\CH = \sum_{i\neq j}^N  
H_{i}\, H_{j}-
2\left( E^2_i \, F^2_{j}+F^2_{i}\, E^2_{j}\right)-
\left(E_i\,F_j -F_i\,E_j \right).
\eeq
Up to a term proportional to the identity, (\ref{ano}) corresponds to the limit 
$z \rightarrow 0$ of the Hamiltonian (\ref{ham}), i.e.
$ \lim_{z \rightarrow 0} \CH_q = \CH + 1/4$.
A complete set of independent commuting observables is provided by
\begin{equation} \label{ossup}
\left\{ \Delta^{(N)}(H),\ C^{(2)},\dots,\ C^{(N)} \right \}. 
\end{equation}
Taking the limit $z \rightarrow 0$ in the definition of the states 
(\ref{stt}--\ref{st}--\ref{vuoto}) (i.e. replacing $\Delta_q$ with $\Delta$)
we obtain the following results:

\begin{prop} The states $\psi(m_l,s_{m_l}, \dots, 0,0)$
are annihilated by $\Delta^{(s_{m_l})}(F)$ iff
\beq
\frac{\alpha_{i+1}}{\alpha_i}  = 
\frac{2\,(-1)^{\delta m -i } \left(\tau +\delta m -i -\half  \right)
-1 - 2 \tau}
{(-1)^{i+1}(1+4j)+2i +1 -4j }, 
\eeq 
where $\tau$ is the eigenvalue of $\Delta^{(s_{m_l})}(H)$.
\end{prop}

\begin{prop} The states $\varphi(k,m_l,s_{m_l}, \dots, 0,0)$
 are eigenvectors of the set (\ref{ossup}), namely
\beq
C^{(n)}\, \varphi(k,m_l,s_{m_l}, \dots, 0,0)
=\l_{n}\, \varphi(k,m_l,s_{m_l}, \dots,0,0),\label{11}
\eeq
with eigenvalues $\l_{n}$ given by
\beq
\l_{n}=
(\rho-i+1)(\rho-i)+ \frac{1}{4}, \label{22}
\eeq
where $\rho$ is the eigenvalue of $\Delta^{(n)}(H)$ on the state $\psi(i,s_i,\dots)$,
and the value of $i \leq l$ is selected by the condition
\beq
s_{m_i} \leq n < s_{m_{i+i}}, \qquad s_{m_{l+1}}=N+1. \label{33}
\eeq
\end{prop}

\subsection{${\CU}_q(\mathfrak{osp}(1 | 2))$ Gaudin model with $j=1/2$}

Here we consider the particular case of the fundamental representation, namely the spin $j=1/2$
one ($1 \leq \delta m \leq 2$). 
This case greatly simplify calculations, allowing a meaningful understanding of the 
results we have presented in the previous section.

Proposition \ref{mostro} becomes the following one.

\begin{prop}The states (\ref{st}) with $\delta m = 1$
are annihilated by $\Delta_q^{(s_{m_l})}(F)$ iff
$$
\alpha_0(z)=1, \qquad
\alpha_1(z)= e^{\frac{z}{2}(\tau-1)}
\frac{\sinh(z \, \tau )}{\sinh z}.
$$
The states (\ref{st}) with $\delta m = 2$
are annihilated by $\Delta_q^{(s_{m_l})}(F)$ iff
$$
\alpha_0(z)=1, \qquad
\alpha_1(z)=-e^{\frac{z \, \tau}{2}}
\frac{\cosh \left[z\left(\tau+\half\right)\right]}
{\cosh \left(\frac{z}{2}\right)}, \qquad 
\alpha_2(z)= e^{z \, \tau } \frac{\sinh ( z \, \tau) \cosh \left[z\left(\tau+\half\right)\right]}
{\sinh z \cosh \left(\frac{z}{2}\right)},
$$
where $\tau=m_{l-1} - s_{m_l} +1 $.

\end{prop}

On the other hand proposition \ref{mostro2} reduces to

\begin{prop} The states (\ref{stt}) are eigenvectors of the set (\ref{oss}), namely
\beq
C^{(n)}(z)\, \varphi_z(k,m_l,s_{m_l}, \dots, 0,0)
=\l_{n}(z)\, \varphi_z(k,m_l,s_{m_l}, \dots,0,0),\nonumber
\eeq
with eigenvalues $\l_{n}(z)$ given by
\beq
\l_{n}(z)=
\frac{\sinh^2 \left[
z\left(n-i + \frac{1}{2}\right)\right]}{\sinh^2z}, \nonumber
\eeq
where the value of $i \leq l$ is selected by the condition
\beq
s_{m_i} \leq n < s_{m_{i+i}}, \qquad s_{m_{l+1}}=N+1. \nonumber
\eeq
\end{prop} 

In this case it is also possible to determine  explicitly the degeneracies of the spectrum. These
obviously correspond to those of the spin $1$ case of the original $\mathfrak {sl}(2)$
Gaudin model. Namely, (\ref{degen}) now reads:
$$
\left[ \CD^{\left(\half \right)}  \right]^{\otimes N} = \bigoplus_{l=0}^{\frac{N}{2}}
c_{\half,l}^{(N)}\,  \CD_l, 
$$
and the following result can be proved by means of the character identity. 
\begin{prop}
The total number of eigenstates $\varphi(k,m_l,s_{m_l}, \dots, 0,0)$ with 
$m_l = N/2 -l $ is given by
$$
 c_{\half,l}^{(N)}= \sum_{k=l}^{\left[\frac{N+l}{2}\right]}\binomial{N}{2k-l}
 \binomial{ 2k-l} { k }-
 \sum_{k=l}^{\left[\frac{N+l-1}{2}\right]}\binomial{N}{2k-l+1}
 \binomial{ 2k-l+1} { k+1} .
$$

\end{prop}

\section{Concluding remarks}

We have constructed a Gaudin model that shares both a deformed  coalgebraic structure and 
a superalgebra symmetry. This has been achieved first deforming the Lie superalgebra  
$ \mathfrak{osp}(1|2) $, endowing it with 
a Hopf structure and then applying to it a slightly modified version of the algorithm
proposed in \cite{BCR,BR,J}.

We obtained an exhaustive description of the spectrum and eigenstates of a particular 
Hamiltonian, wich reduces to the known one for the Gaudin model associated to 
$ \mathfrak{osp}(1|2) $ \cite{J,KU}.

Our approach, thanks to its purely algebraic nature, can be obviously used 
for any spin of the representation.

\section*{Acknowledgements}

One of us (GS) would like to thank D. Arnaudon, L. Frappat and E. Ragoucy for interesting discussions.
The hospitality of LAPTH, extended to GS during his
visit to Annecy in 2004, where part of this work was done, is  also kindly acknowledged.

\section*{Appendix 1: Some remarks about $\mathfrak{osp}(1|2)$ representations }

In this appendix we shall recall the basic concepts of graded vector
spaces as modules of superalgebras representations. 
Graded vector spaces are vector spaces equipped with a notion of odd
and even degree, that allows us to treat fermions. 

Let $\rho$ be an irreducible finite--dimensional representation of the Lie superalgebra 
$\mathfrak{osp}(1|2)$, $\rho: \mathfrak{osp}(1|2) \mapsto {\rm{End}}(V)$, where $V$ is the
module of the representation.

The following results hold \cite{FR}: 

\begin{itemize}

\item ${\rm{dim}}(V)=4j+1$, where $j$ is a non negative integer or half integer, called
the {\it{spin}} of the representation $\rho$;

\item $V = V_0 \oplus V_1$ with  $\dim V_0 = 2j+1$ and 
$\dim V_1 = 2j$. We shall call $v_0 \in V_0$ even (or bosonic) and $v_1 \in V_1$
odd (or fermionic). The subspaces $V_0$ and $V_1$ are called the homogeneous components
of $V$. If we choose a basis of homogeneous elements $e_i \in V$, $i=0,1,\dots, 4j+1$, we can define 
a grading $\CG: i \rightarrow \mathbb{Z}_2$:
$$
\CG(i) \doteq
\left\{\begin{array}{cc}
0 & {\rm{if}} \; e_i \in V_0, \\
1 & {\rm{if}} \; e_i \in V_1.
\end{array}\right.
$$

\item Elements $[ \rho (a)]_{ik}$, $a \in \mathfrak{osp}(1|2)$, $i,k =1,...,4j+1$ of
the representation $\rho$
have a grading $\pi: \{1,...,4j+1\} \times \{1,...,4j+1\}  \rightarrow \mathbb{Z}_2$
such that
$$
\pi [ \rho (a)]_{ik} \doteq  \CG(i) + \CG (k)   \qquad {\rm{mod}}\,2.
$$
The bosonic (resp. fermionic) sector is the set of elements with $\pi =0$ (resp. $\pi =1$).

In this way we give a graded structure (that we call ``grading'')
both to the module $V$ of the representation
and to the elements of ${\rm{End}}(V)$, thus reflecting the ``gradation'' of
the elements of $\mathfrak{osp}(1|2)$.

\item 
The bosonic sector of each representation $\rho_j$ is the completely reducible
representation of $\mathfrak{sl}(2)$ given by
$$
\rho_j = \CD_j \oplus \CD_{j - 1/2}, \qquad j \neq 0.
$$

\item 
The tensor product of two irreducible representations 
$\rho_{j}$ and  $\rho_{k}$ is given by:
$$
\rho_{j} \otimes \rho_{k} = \bigoplus_{J= | j -k |}^{j+k} \rho_J,
$$
whit $J$ integer or half-integer.

\end{itemize}

Let $v_i \in V$ and 
$A_i \, \in {\rm{End}}(V)$, $i=1,...,N$ be respectively $N$ homogeneous vectors
and $N$ homogeneous endomorphisms. Hence we can construct the operator
$A_1 \otimes \cdots \otimes A_N \in {\rm{End}}(V)^{\otimes N}$ and the vector
$v_1 \otimes \cdots \otimes v_N \in V^{\otimes N}$ using (\ref{bg}).

One can check that it holds the following

\begin{prop} 
The action of $A_1 \otimes \cdots \otimes A_N \in {\rm{End}}(V)^{\otimes N}$
on $v_1 \otimes \cdots \otimes v_N \in V^{\otimes N}$ is associative if
$$
(A_1 \otimes \cdots \otimes A_N )\, (v_1 \otimes \cdots \otimes v_N) =
(-1)^{\sum_{i=2}^N {\rm{deg}}(A_i)  \sum_{j=1}^{i-1} \CG (v_j) }
(A_1 \, v_1) \otimes \cdots \otimes (A_N \, v_N).
$$
\end{prop}

The above proposition is a key point in applying the algorithm (both in the
deformed case and in the non-deformed case) exposed in Section 3.

\section*{Appendix 2: Co-existence of two families of eigenstates}

Let us consider the ${\CU}_q(\mathfrak{osp}(1 | 2))$ Gaudin model with $j=1/2$ and $N=2$
in order to show the occurence of two complete families of eigenstates, corresponding
to the possible choices ${\rm{deg}} \ket{\downarrow}=1$ (fermion--boson--fermion) and 
${\rm{deg}}\ket{\downarrow}=0$ (boson--fermion--boson).

In the first case we obtain the following results: 
$$
\begin{tabular}{|l|p{10 cm}|p{1.8 cm}|}
\hline
$fbf$        &  $\varphi(k;m_l,s_{m_l};\cdots,0,0)\ $ &   $C^{(2)}$         \\ 
\hline
\hline
             & $\varphi(0;0,0)=\ket{\downarrow\downarrow}$ &   \\
\cline{2-2} 
             & $\varphi(1;0,0)=q^{-1/2}\ket{0 \downarrow}-q^{1/2}\ket{\downarrow 0}$ &  \\
\cline{2-2} 
 $\psi(0,0)$ & $\varphi(2;0,0)=q^{-1}\ket{\uparrow \downarrow}-(q^{1/2}-q^{-1/2})\ket{00}+
              q\ket{\downarrow \uparrow}$ &  $ \frac{\sinh^2(5z/2)}
              {\sinh^2 z} $   \\ 
\cline{2-2}   
             & $\varphi(3;0,0)=q^{-3/2}\ket{\uparrow 0}+q^{3/2}\ket{0 \uparrow}     
                +(q^{1/2}-q^{-1/2})(\ket{\uparrow 0}+\ket{0 \uparrow})          $ &        \\
\cline{2-2}  
             & $\varphi(4;0,0)=\ket{\uparrow\uparrow}$ &   \\ 
\hline
             & $\varphi(1;1,2)= q^{1/2}\ket{0 \downarrow}+q^{-1/2}\ket{\downarrow 0}$ &        \\
\cline{2-2}
 $\psi(1,2)$ & $\varphi(2;1,2)= \ket{\uparrow \downarrow}-(q^{1/2}+q^{-1/2})\ket{00} 
               -\ket{\downarrow \uparrow}$  & $ \frac{\sinh^2(3z/2)} 
               {\sinh^2 z} $   \\ 
\cline{2-2}
             & $\varphi(3;1,2)= q^{1/2}\ket{\uparrow 0}+q^{-1/2}\ket{0 \uparrow}$ &         \\
\hline  

 $\psi(2,2)$ & $\varphi(2;2,2)= q^{1/2}\ket{\uparrow \downarrow}+\ket{00} 
               -q^{-1/2}\ket{\downarrow \uparrow}$  & $ \frac{\sinh^2(z/2)} 
               {\sinh^2 z} $   \\
\hline                   
 \end{tabular}
$$
while the second choice leads to
$$
\begin{tabular}{|l|p{10 cm}|p{1.8 cm}|}
\hline
$bfb$        &  $\varphi(k;m_l,s_{m_l};\cdots,0,0)\ $ &   $C^{(2)}$         \\ 
\hline
\hline
             & $\varphi(0;0,0)=\ket{\downarrow\downarrow}$ &   \\
\cline{2-2} 
             & $\varphi(1;0,0)=q^{-1/2}\ket{0 \downarrow}+q^{1/2}\ket{\downarrow 0}$ &  \\
\cline{2-2} 
 $\psi(0,0)$ & $\varphi(2;0,0)=q^{-1}\ket{\uparrow \downarrow}+(q^{1/2}-q^{-1/2})\ket{00}+
              q\ket{\downarrow \uparrow}$ &  $ \frac{\sinh^2(5z/2)}
              {\sinh^2 z} $   \\ 
\cline{2-2}   
             & $\varphi(3;0,0)=q^{-3/2}\ket{\uparrow 0}-q^{3/2}\ket{0 \uparrow}
             -(q^{1/2}-q^{-1/2})(\ket{\uparrow 0}+\ket{0 \uparrow})          $ &        \\
\cline{2-2}  
             & $\varphi(4;0,0)=\ket{\uparrow\uparrow}$ &   \\ 
\hline
             & $\varphi(1;1,2)= q^{1/2}\ket{0 \downarrow}-q^{-1/2}\ket{\downarrow 0}$ &        \\
\cline{2-2}
 $\psi(1,2)$ & $\varphi(2;1,2)= \ket{\uparrow \downarrow}+(q^{1/2}+q^{-1/2})\ket{00} 
               -\ket{\downarrow \uparrow}$  & $ \frac{\sinh^2(3z/2)} 
               {\sinh^2 z} $   \\ 
\cline{2-2}
             & $\varphi(3;1,2)= q^{1/2}\ket{\uparrow 0}-q^{-1/2}\ket{0 \uparrow}$ &         \\
\hline  
 $\psi(2,2)$ & $\varphi(2;2,2)= q^{1/2}\ket{\uparrow \downarrow}-\ket{00} 
               -q^{-1/2}\ket{\downarrow \uparrow}$  & $ \frac{\sinh^2(z/2)} 
               {\sinh^2 z} $   \\
\hline                   
 \end{tabular}
 $$

Let us notice that the eigenvalues of $C^{(2)}$ and their degeneracies are the same in both cases.


\end{document}